\newif\ifDRAFT
\newif\ifBELLENOTE
\newif\ifBELLEPREPRINT
\newif\ifBELLE
\newif\ifHEPEX
\newif\ifPULL
\BELLEPREPRINTtrue
\PULLtrue
\ifBELLENOTE\BELLEtrue\fi
\ifBELLEPREPRINT\BELLEtrue\fi
\ifBELLE
\documentclass[floatfix,preprint,superscriptaddress,showpacs,preprintnumbers,amsmath,amssymb]{revtex4-1}
\else
\documentclass[aps,prl,floatfix,twocolumn,superscriptaddress,showpacs,preprintnumbers,amsmath,amssymb]{revtex4-1}
\fi
\usepackage{graphicx} 
\usepackage{dcolumn}  
\ifDRAFT
\usepackage{lineno}   
\fi
\usepackage{xspace}   

\graphicspath{{figs}}

\def\ee     {\ensuremath{e^+e^-}\xspace}
\def\mumu   {\ensuremath{\mu^+\mu^-}\xspace}
\def\tautau {\ensuremath{\tau^+\tau^-}\xspace}
\def\fb     {\ensuremath{\,\mathrm{fb}^{-1}}\xspace}

\def\MeV    {\ensuremath{\,\mathrm{MeV}}\xspace}

\def\MeVcc  {\ensuremath{\,\mathrm{MeV}/c^2}\xspace}
\def\GeV    {\ensuremath{\,\mathrm{GeV}}\xspace}

\def\GeVcc  {\ensuremath{\,\mathrm{GeV}/c^2}\xspace}
\def\mum    {\ensuremath{\,\mu\mathrm{m}}\xspace}

\def\cm     {\ensuremath{\,\mathrm{cm}}\xspace}
\def\mrad   {\ensuremath{\,\mathrm{mrad}}\xspace}
\def\s      {\ensuremath{\,\mathrm{s}}\xspace}
\def\stat   {\ensuremath{\mathrm{stat.}}\xspace}
\def\syst   {\ensuremath{\mathrm{syst.}}\xspace}
\def\BF     {\ensuremath{{\cal B}}\xspace}
\def\Ebeam  {\ensuremath{E^{*}_\mathrm{beam}}\xspace}

\newcommand{\qq}[1]{\ensuremath{#1\bar{#1}}\xspace}
\newcommand{\mean}[1]{\ensuremath{\langle{#1}\rangle}\xspace}
\def\ct       {\ensuremath{t}\xspace}
\def\ctau     {\ensuremath{\tau}\xspace}
\def\meanctau {\ensuremath{\langle{\tau}\rangle}\xspace}
\def\dct      {\ensuremath{\Delta t}\xspace}

\begin{document}
\newlength{\figwidth}
\setlength{\figwidth}{0.85\columnwidth}
\ifBELLE
\setlength{\figwidth}{0.80\columnwidth}
\fi
%
\ifBELLE
\vspace*{-1.5cm}
\resizebox{!}{3cm}{\includegraphics{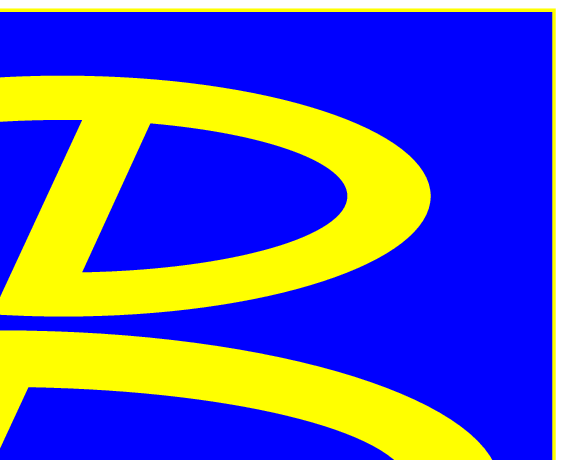}}
\vspace*{-1.5cm}
\fi
\preprint{\vbox{\ifDRAFT
\else
                 \ifBELLEPREPRINT
                  \hbox{KEK Preprint 2013-44}
                  \hbox{Belle Preprint 2013-26}
                  \hbox{Submitted to {\it PRL}}
  		          \ifHEPEX\hbox{arXiv: XXXX.YYYY [hep-ex]}\fi
                  \hbox{\today}
                 \fi
                \fi
}}
\title{\boldmath\quad\\[1.0cm] Measurement of the $\tau$-lepton
lifetime at Belle}
%
\noaffiliation
\affiliation{University of the Basque Country UPV/EHU, 48080 Bilbao}
\affiliation{Beihang University, Beijing 100191}
\affiliation{Budker Institute of Nuclear Physics SB RAS and Novosibirsk State University, Novosibirsk 630090}
\affiliation{Faculty of Mathematics and Physics, Charles University, 121 16 Prague}
\affiliation{University of Cincinnati, Cincinnati, Ohio 45221}
\affiliation{Deutsches Elektronen--Synchrotron, 22607 Hamburg}
\affiliation{Justus-Liebig-Universit\"at Gie\ss{}en, 35392 Gie\ss{}en}
\affiliation{Hanyang University, Seoul 133-791}
\affiliation{University of Hawaii, Honolulu, Hawaii 96822}
\affiliation{High Energy Accelerator Research Organization (KEK), Tsukuba 305-0801}
\affiliation{Hiroshima Institute of Technology, Hiroshima 731-5193}
\affiliation{Ikerbasque, 48011 Bilbao}
\affiliation{Indian Institute of Technology Guwahati, Assam 781039}
\affiliation{Indian Institute of Technology Madras, Chennai 600036}
\affiliation{Institute of High Energy Physics, Chinese Academy of Sciences, Beijing 100049}
\affiliation{Institute of High Energy Physics, Vienna 1050}
\affiliation{Institute for High Energy Physics, Protvino 142281}
\affiliation{INFN - Sezione di Torino, 10125 Torino}
\affiliation{Institute for Theoretical and Experimental Physics, Moscow 117218}
\affiliation{J. Stefan Institute, 1000 Ljubljana}
\affiliation{Kanagawa University, Yokohama 221-8686}
\affiliation{Institut f\"ur Experimentelle Kernphysik, Karlsruher Institut f\"ur Technologie, 76131 Karlsruhe}
\affiliation{Korea Institute of Science and Technology Information, Daejeon 305-806}
\affiliation{Korea University, Seoul 136-713}
\affiliation{Kyungpook National University, Daegu 702-701}
\affiliation{\'Ecole Polytechnique F\'ed\'erale de Lausanne (EPFL), Lausanne 1015}
\affiliation{Faculty of Mathematics and Physics, University of Ljubljana, 1000 Ljubljana}
\affiliation{Luther College, Decorah, Iowa 52101}
\affiliation{University of Maribor, 2000 Maribor}
\affiliation{Max-Planck-Institut f\"ur Physik, 80805 M\"unchen}
\affiliation{School of Physics, University of Melbourne, Victoria 3010}
\affiliation{Moscow Physical Engineering Institute, Moscow 115409}
\affiliation{Moscow Institute of Physics and Technology, Moscow Region 141700}
\affiliation{Graduate School of Science, Nagoya University, Nagoya 464-8602}
\affiliation{Kobayashi-Maskawa Institute, Nagoya University, Nagoya 464-8602}
\affiliation{Nara Women's University, Nara 630-8506}
\affiliation{National Central University, Chung-li 32054}
\affiliation{National United University, Miao Li 36003}
\affiliation{Department of Physics, National Taiwan University, Taipei 10617}
\affiliation{H. Niewodniczanski Institute of Nuclear Physics, Krakow 31-342}
\affiliation{Nippon Dental University, Niigata 951-8580}
\affiliation{Niigata University, Niigata 950-2181}
\affiliation{University of Nova Gorica, 5000 Nova Gorica}
\affiliation{Osaka City University, Osaka 558-8585}
\affiliation{Pacific Northwest National Laboratory, Richland, Washington 99352}
\affiliation{Panjab University, Chandigarh 160014}
\affiliation{University of Pittsburgh, Pittsburgh, Pennsylvania 15260}
\affiliation{University of Science and Technology of China, Hefei 230026}
\affiliation{Seoul National University, Seoul 151-742}
\affiliation{Soongsil University, Seoul 156-743}
\affiliation{Sungkyunkwan University, Suwon 440-746}
\affiliation{School of Physics, University of Sydney, NSW 2006}
\affiliation{Tata Institute of Fundamental Research, Mumbai 400005}
\affiliation{Excellence Cluster Universe, Technische Universit\"at M\"unchen, 85748 Garching}
\affiliation{Toho University, Funabashi 274-8510}
\affiliation{Tohoku Gakuin University, Tagajo 985-8537}
\affiliation{Tohoku University, Sendai 980-8578}
\affiliation{Department of Physics, University of Tokyo, Tokyo 113-0033}
\affiliation{Tokyo Institute of Technology, Tokyo 152-8550}
\affiliation{Tokyo Metropolitan University, Tokyo 192-0397}
\affiliation{Tokyo University of Agriculture and Technology, Tokyo 184-8588}
\affiliation{University of Torino, 10124 Torino}
\affiliation{CNP, Virginia Polytechnic Institute and State University, Blacksburg, Virginia 24061}
\affiliation{Wayne State University, Detroit, Michigan 48202}
\affiliation{Yamagata University, Yamagata 990-8560}
\affiliation{Yonsei University, Seoul 120-749}
  \author{K.~Belous}\affiliation{Institute for High Energy Physics, Protvino 142281} 
  \author{M.~Shapkin}\affiliation{Institute for High Energy Physics, Protvino 142281} 
  \author{A.~Sokolov}\affiliation{Institute for High Energy Physics, Protvino 142281} 
  \author{I.~Adachi}\affiliation{High Energy Accelerator Research Organization (KEK), Tsukuba 305-0801} 
  \author{H.~Aihara}\affiliation{Department of Physics, University of Tokyo, Tokyo 113-0033} 
  \author{D.~M.~Asner}\affiliation{Pacific Northwest National Laboratory, Richland, Washington 99352} 
  \author{V.~Aulchenko}\affiliation{Budker Institute of Nuclear Physics SB RAS and Novosibirsk State University, Novosibirsk 630090} 
  \author{A.~M.~Bakich}\affiliation{School of Physics, University of Sydney, NSW 2006} 
  \author{A.~Bala}\affiliation{Panjab University, Chandigarh 160014} 
  \author{B.~Bhuyan}\affiliation{Indian Institute of Technology Guwahati, Assam 781039} 
% ...
\collaboration{The Belle Collaboration}

\noaffiliation

\begin{abstract}
The lifetime of the $\tau$-lepton is measured using the process $\ee
\rightarrow\tautau$, where both $\tau$-leptons decay to $3\pi\nu_\tau$.
The result for the mean lifetime, based on $711\fb$ of data collected
with the Belle detector at the $\Upsilon(4S)$ resonance and $60\MeV$
below, is $\tau = (290.17 \pm 0.53(\stat) \pm 0.33(\syst))
\cdot10^{-15}\s$. The first measurement of the lifetime difference
between $\tau^+$ and $\tau^-$ is performed. The upper limit on the
relative lifetime difference between positive and negative
$\tau$-leptons is $|\Delta\tau| / \tau < 7.0 \times 10^{-3}$ at 90\% CL.
\end{abstract}


\maketitle

\ifDRAFT
\else
\tighten
\fi

{\renewcommand{\thefootnote}{\fnsymbol{footnote}}}
\setcounter{footnote}{0}

\ifBELLE
\section{Introduction}
\fi

High precision measurements of the mass, lifetime and leptonic branching
fractions of the $\tau$-lepton can be used to test lepton
universality~\cite{LU}, which is assumed in the Standard Model.
Among the recent experimental results that may manifest the violation of
the lepton universality in the case of the $\tau$-lepton, the combined
measurement  of the ratio of the branching fraction of $W$-boson decay
to $\tau\nu_\tau$ to the mean branching fraction of $W$-boson decay to
$\mu\nu_\mu$ and $e\nu_e$ by the four LEP experiments stands out:
$2\BF(W\rightarrow\tau\nu_\tau) / (\BF(W\rightarrow\mu\nu_\mu) +
\BF(W\rightarrow e\nu_e)) = 1.066 \pm 0.025$~\cite{LEPEW}, which
differs from unity by 2.6 standard deviations. The present PDG value
of the $\tau$-lepton lifetime $(290.6\pm1.0)\cdot10^{-15}\s$~\cite{PDG}
is dominated by the results obtained in the LEP experiments~\cite{LEP}.

A high-statistics data sample collected at Belle allows us to select
\tautau events where both $\tau$-leptons decay to three charged pions
and a neutrino. As explained later, for these events the directions of
the $\tau$-leptons can be determined with an accuracy better than that
given by the thrust axis of the event. At an asymmetric-energy collider,
the laboratory frame angle between the produced $\tau$-leptons is not
equal to 180 degrees, so their production point can be determined from
the intersection of two trajectories defined by the $\tau$-lepton decay
vertices and their momentum directions. The direction of each
$\tau$-lepton in the laboratory system can be determined with twofold
ambiguity. These special features of the asymmetric-energy B-factory
experiments allow a high precision measurement of the $\tau$-lepton
lifetime with systematic uncertainties that differ from those of the
LEP experiments.

Furthermore, Belle's asymmetric-energy collisions provide a unique
possibility to measure separately the $\tau^+$ and $\tau^-$ lifetimes,
which allows us to test $CPT$ symmetry in $\tau$-lepton decays.

\ifBELLE
\section{Description of the measurement method and selection criteria}
\else
~
\fi

In the following, we use symbols with and without an asterisk for
quantities in the \ee center-of-mass (CM) and laboratory frame,
respectively. In the CM frame, $\tau^+$ and $\tau^-$ leptons emerge back
to back with the energy $E^*_\tau$ equal to the beam energy \Ebeam if we
neglect the initial- (ISR) and final-state radiation (FSR). We determine
the direction of the $\tau$-lepton momentum in the CM frame as follows.
If the neutrino mass is assumed to be zero for the hadronic decay
$\tau\rightarrow X\nu_\tau$ ($X$ representing the hadronic system with
mass $m_X$ and energy $E^*_X$), the angle $\theta^*$ between the
momentum $\vec{P}^*_X$ of the hadronic system and that of the
$\tau$-lepton is given by:
\begin{equation}
\cos\theta^* = \frac{2E^*_\tau E^*_X-m_\tau^2-m_X^2}
                     {2P^*_X\sqrt{(E^*_\tau)^2-m_\tau^2}}.
\label{eq:cost}
\end{equation}
The requirement that the $\tau$-leptons be back to back in the CM can be
written as a system of three equations: two linear and one quadratic.
For the components $x^*$, $y^*$, $z^*$ of the unit vector $\hat{n}^*_+$
representing the direction of the positive $\tau$-lepton, we write:
\begin{equation}
\left\{ \begin{array}{l}
x^*\cdot P^*_{1x}+y^*\cdot P^*_{1y}+z^*\cdot P^*_{1z}= |P^*_1|\cos\theta^*_1 \\
x^*\cdot P^*_{2x}+y^*\cdot P^*_{2y}+z^*\cdot P^*_{2z}=-|P^*_2|\cos\theta^*_2 \\
(x^*)^2+(y^*)^2+(z^*)^2=1
\label{eq:sys}
\end{array} \right.
\end{equation}
where $\vec{P}^*_1$ and $\vec{P}^*_2$ are the momenta of the hadronic
systems in the CM and $\cos\theta^*_i$ ($i=1,2$) are given by
Eq.~\eqref{eq:cost}. Index 1 (2) is used for the positive (negative)
$\tau$-lepton.

There are two solutions for Eq.~\eqref{eq:sys}, so the direction
$\hat{n}^*_+$ is determined with twofold ambiguity. In the present
analysis, we take the mean vector of the two solutions of
Eq.~\eqref{eq:sys} as the direction of the $\tau$-lepton in CM. The
analysis of MC simulated events shows that there is no bias due to this
choice.

\begin{figure}[htb]
\vspace*{-0.5cm}
\includegraphics[width=0.65\figwidth]{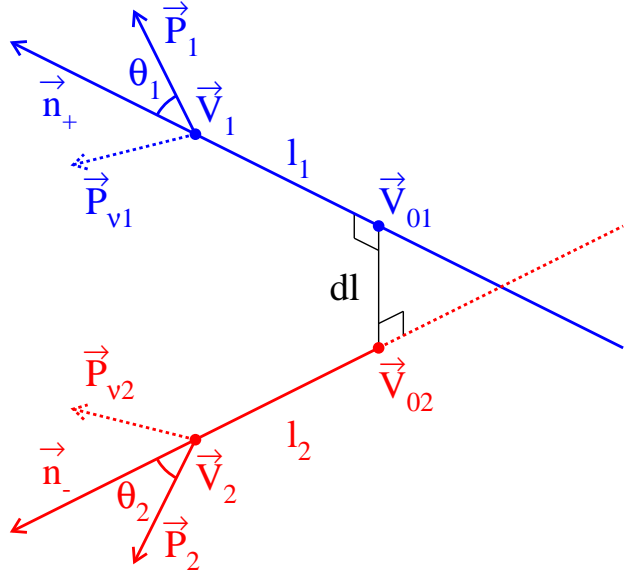}
\vspace*{-0.5cm}
\caption{\label{fg:fig2}The schematic view of the \tautau event in the
laboratory frame.}
\end{figure}

Each direction $\hat{n}_{\pm}^*$ is converted to a four-momentum using
the $e^{\pm}$ beam energy and the $\tau$ mass. Both four-momenta are
then boosted into the laboratory frame, each passing through the
corresponding $\tau$ decay vertex $\vec{V}_i$ that is determined by the
three pion-daughter tracks (see Fig.~\ref{fg:fig2}). We approximate the
trajectory of $\tau$-leptons in the magnetic field of the Belle detector
with a straight line. Due to the finite detector resolution, these
straight lines do not intersect at the \tautau production point. The
three-dimensional separation between these lines is characterized by the
distance $dl$ between the two points ($\vec{V}_{01}$ and $\vec{V}_{02}$)
of closest approach. The typical size of $dl$ is $\sim 0.01\cm$. For the
production point of each $\tau$-lepton, we take the points
$\vec{V}_{01}$ and $\vec{V}_{02}$. The flight distance $l_1$ ($l_2$) of
the $\tau^+$ ($\tau^-$) in the laboratory frame is defined as the
distance between the points $\vec{V}_1$ and $\vec{V}_{01}$ ($\vec{V}_2$
and $\vec{V}_{02}$). The proper time \ct (the product of the speed of
light and the decay time of $\tau$-lepton) for the positive
$\tau$-lepton is equal to the distance $l_1$ divided by its relativistic
kinematic factor $\beta\gamma$ in the laboratory frame: $\ct_1 =
l_1/(\beta\gamma)_1$. The corresponding parameter for the negative
$\tau$-lepton is $\ct_2 = l_2/(\beta\gamma)_2$.

The analysis presented here is based on the data collected with the
Belle detector~\cite{Belle} at the KEKB asymmetric-energy \ee ($3.5$ on
$8\GeV$) collider~\cite{KEKB} operating at the $\Upsilon(4S)$ resonance
and $60\MeV$ below. The total integrated luminosity of the data sample
used in the analysis is $711\fb$. Two inner detector configurations were
used. A $2.0\cm$ beampipe and a 3-layer silicon vertex detector (SVD1)
were used for the first sample of $157\fb$, while a 1.5 cm beampipe, a
4-layer silicon detector (SVD2) and a small-cell inner drift chamber
were used to record the remaining $554\fb$~\cite{SVD2}. The integrated
luminosity of the data sample at the energy below the $\Upsilon$(4S)
resonance is about 10\% of the total data sample. All analyzed
distributions for the on- and off-resonance data coincide within the
statistical uncertainties with each other; this justifies our
combination of the on- and off-resonance \ct distributions in the
present analysis.

The following requirements are applied for the selection of the \tautau
events where both $\tau$-leptons decay into three charged pions and a
neutrino:

1.  there are exactly six charged pions with zero net charge and there
    are no other charged tracks;

2.  the $K^0_S$ mesons, $\Lambda$-hyperons and $\pi^0$ are found by
    $V^0$~\cite{V0} and $\pi^0$ reconstruction algorithms and the event
    is discarded if any of these are seen;

3.  the number of photons should be smaller than six and their total
    energy should be less than $0.7\GeV$;

4.  the thrust value of the event in the CM frame is greater than 0.9;

5.  the square of the transverse momentum of the $6\pi$ system is
    required to be greater than $0.25\,(\mathrm{GeV}/c)^2$ to suppress
    the $\ee\rightarrow\ee 6\pi$ two-photon events;

6.  the mass $M(6\pi)$ of the $6\pi$ system should fulfill the
    requirement $4\GeVcc<M(6\pi)<10.25\GeVcc$ to suppress other
    background events;

7.  there should be three pions (triplet) with net charge equal to
    $\pm1$ in each hemisphere (separated by the plane perpendicular to
    the thrust axis in the CM);

8.  the pseudomass (see the definition in Ref.~\cite{taumass}) of each
    triplet of pions must be less than $1.8\GeVcc$;

9.  each pion-triplet vertex-fit quality must satisfy $\chi^2<20$;

10. the discriminant $D$ of Eq.~\eqref{eq:sys} should satisfy $D>-0.05$
    (with slightly negative values arising from experimental
    uncertainties; if this happens, we use $D=0$ when
    solving the equation);

11. the distance of closest approach  must satisfy $dl<0.03\cm$ to
    reject events with large uncertainties in the reconstructed momenta
    and vertex positions.

All of these selection criteria are based on a study of the signal and
background Monte Carlo (MC) simulated events.

For the signal MC sample, we use \tautau events produced by the KKMC
generator~\cite{KKMC} with the mean lifetime $\meanctau = 87.11\mum$
that are then fed to the full detector simulation based on
GEANT~3~\cite{GEANT}. These events are passed through the same
reconstruction procedures as for the data.

For the background estimation, we use the MC samples of events generated
by the EVTGEN program~\cite{EVTGEN}, which correspond to the one-photon
annihilation diagram  $\ee\rightarrow\qq{q}$, where $\qq{q}$ are
$\qq{u}$, $\qq{d}$, $\qq{s}$ ($uds$ events), $\qq{c}$ (charm events),
and $\ee\rightarrow\Upsilon(4S)\rightarrow B^+B^-, B^0\bar{B^0}$ (beauty
events). All these events are passed through the full detector
simulation and reconstruction procedures. The statistics in these MC
samples are equivalent to the integrated luminosity of the data,
\textit{i.e.,} the number of events of a given category is equal to the
product of the integrated luminosity of the data and the expected cross
section from theory. For the estimation of the background from the
process $\gamma\gamma \rightarrow hadrons$ ($\gamma\gamma$ events), we
use events generated by PYTHIA~\cite{PYTHIA} that are subjected to the
afore-mentioned simulation and reconstruction procedures.

In addition to the above MC samples, we also use two
$\ee\rightarrow\tautau$ MC samples, generated by KKMC, where both
$\tau$-leptons are forced to decay into three charged pions and a
neutrino. The mean lifetimes for these two samples are $84$ and
$90\mum$, which are about $10\sigma$ below and above the PDG value.
These two samples are also passed through the same detector simulation
and reconstruction procedures.

\ifBELLE
\section{Analysis of the experimental results}
\else
~
\fi

\begin{figure}[htb]
\vspace*{-0.2cm}
\includegraphics[width=\figwidth]{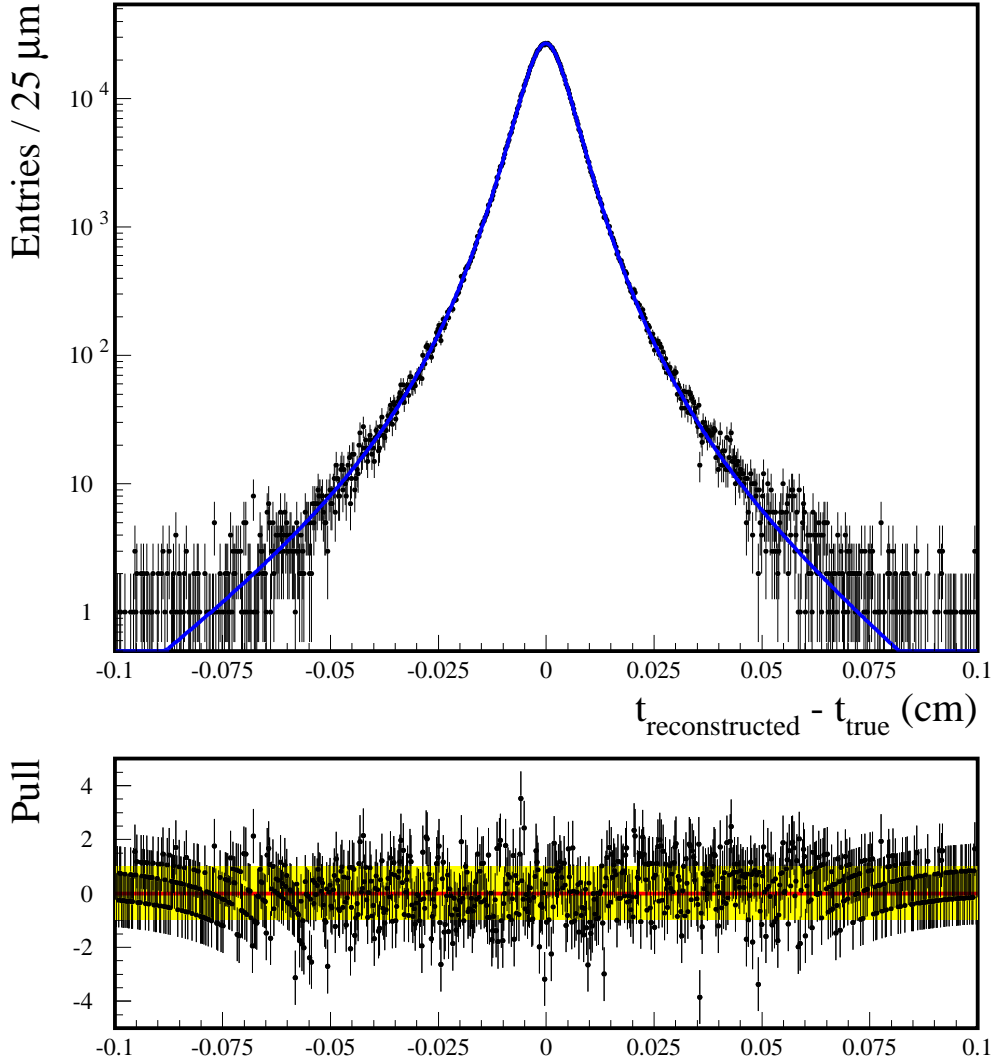}
\caption{\label{fg:res0}Distribution of the difference between the
reconstructed and true \ct values for $\tau$-leptons (obtained from an
MC sample) for the combined SVD1- and SVD2-geometry data sets. The line
is the result of the fit to Eq.~\eqref{eq:res}.
The distribution of residuals [(data--fit)/error] for the fit is shown
in the bottom panel.}
\end{figure}

In the measured proper time distribution, the exponential behavior is
smeared by the experimental resolution. This resolution has been studied
with MC simulation. The following samples are used, each one with a
slightly different time resolution: with the SVD1 and SVD2 geometries
and three different values of the mean $\tau$-lepton lifetime. For all
MC samples, the resolution function is found to be described well by the
expression:
\begin{equation}\begin{array}{l}
R(\dct) = (1 - A\dct) e^{-(\dct-\ct_0)^2/2\sigma^2}\mathrm{, where}\\
\dct = \ct_\mathrm{reconstructed} - \ct_\mathrm{true},\quad
\dct_0 = \dct - \ct_0,\\
\sigma = a + b|\dct_0|^{1/2} + c|\dct_0| + d|\dct_0|^{3/2}
\label{eq:res}
\end{array}\end{equation}
The parameters $\ct_0$, $a$, $b$, $c$ and $d$ are allowed to vary freely
in the fit, while the asymmetry $A=2.5\cm^{-1}$ is fixed because of its
strong correlation with the lifetime parameter \ctau. An example of the
fitting of the resolution distribution for the MC sample with mean
$\tau$-lepton lifetime equal to $87.11\mum$ and for the sum of the SVD1-
and SVD2-geometry data sets by the function Eq.~\eqref{eq:res} is shown
in Fig.~\ref{fg:res0}. The goodness of fit is $\chi^2/ndf=770.8/794$.
In an alternate fit where the parameter $A$ is allowed to vary freely,
its best-fit value is $(2.5 \pm 0.2)\cm^{-1}$. All of the other
resolution distributions are described with the same level of quality.

After applying all the selection criteria, the contamination of the
background in the data is about two percent. The dominant background
arises from $uds$ events. For these events, all six pions emerge
(typically) from one primary vertex and these $uds$ events are similar
to the \tautau events with zero lifetime. Using the MC, we check that
the decay time distributions of $uds$-events that pass the selection
criteria are well described by the resolution function of
Eq.~\eqref{eq:res}. The same behavior is found for $\gamma\gamma$
events, whose fraction in all the selected events is about
$1.4\cdot 10^{-4}$. Other sources of background contribute to the
selected data sample at the per mille level.

The measured proper time distribution is parameterized by:
\ifBELLE
\begin{equation}
F(\ct) = N \int{e^{-\ct'/\ctau} R(\ct-\ct') d\ct'}
       + A_{uds}R(\ct) + B_{cb}(\ct),
\label{eq:fitrd}
\end{equation}
\else
\begin{equation}\begin{array}{r}
F(\ct) = N \int{e^{-\ct'/\ctau} R(\ct-\ct') d\ct'}\\
       + A_{uds}R(\ct) + B_{cb}(\ct),
\label{eq:fitrd}
\end{array}\end{equation}
\fi
where the resolution function $R(\ct)$ is given by Eq.~\eqref{eq:res},
$A_{uds}$ is the normalization of the combined $uds$ and $\gamma\gamma$
background and $B_{cb}(\ct)$ is the background distribution due to
charm and beauty events. The shapes and yields of the backgrounds
($B_{cb}(\ct)$, $A_{uds}$) are fixed from the MC simulation; the free
parameters of the fit are the normalization $N$, the $\tau$-lepton
lifetime $\ctau$ and the five parameters of the resolution function
$\ct_0$, $a$, $b$, $c$ and $d$.

\begin{figure}[htb] 
\vspace*{-0.2cm}
\includegraphics[width=\figwidth]{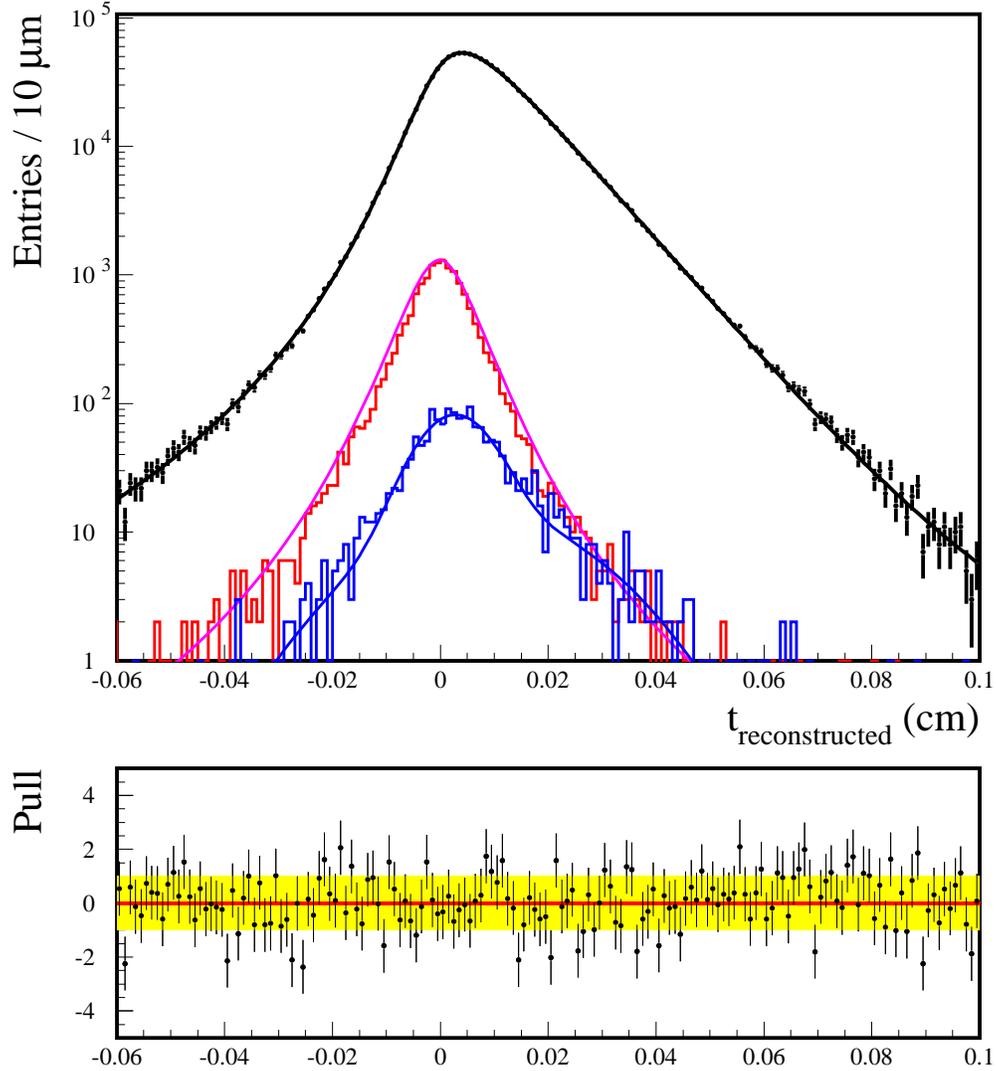}
\caption{\label{fg:ctr}The measured proper time \ct
distribution for the data (filled circles with errors). The black line
is the result of the fit by Eq.~\eqref{eq:fitrd}. The red histogram is
the MC prediction for the sum of the $uds$ and $\gamma\gamma$ background
contributions. The magenta line is the contribution for $uds+\gamma
\gamma$ obtained in the fit. The blue histogram is the MC prediction for
the sum of the charm and beauty background contributions. The blue line
is the smoothed distribution of the charm and beauty contributions that
is used in the fit. The distribution of residuals [(data--fit)/error]
for the fit is shown in the bottom panel.}
\end{figure}

The result of the fit of the experimental data to Eq.~\eqref{eq:fitrd}
is shown in Fig.~\ref{fg:ctr}, together with the contributions from the
sum of $uds$ and $\gamma\gamma$ events and the sum of charm and beauty
events. The curves on these contributions are the result of the fit with
Eq.~\eqref{eq:fitrd}, $A_{uds}R(\ct)$ function (with fixed value of
$A_{uds}$) for $uds$ plus $\gamma\gamma$ events and the fixed sum of two
Gaussians for charm plus beauty events.

The relation of the parameter \ctau in Eq.~\eqref{eq:fitrd} to the
generated value of the $\tau$-lepton mean lifetime is analyzed using
three MC \tautau samples with the mean lifetime values of $84$, $87.11$
and $90\mum$. The dependence of parameter \ctau on the input mean
lifetime value \meanctau is found to be linear; $(\ctau - 87) =
(0.97\pm0.03)(\meanctau - 87) + (0.001\pm0.07)$ [in units of \mum] with
$\chi^2$ of 0.2.

Bias arising from the selection criteria is checked by fitting the
proper time distribution for the signal MC sample before and after
applying cuts, and no bias is found for all the selection criteria
listed above. To check that the fitting procedure gives the correct
estimation of the input lifetime value for different resolution
functions, we perform the fits of the decay time distributions for MC
samples with a lifetime of $87.11\mum$ for the sum of the SVD1 and SVD2
samples, SVD1 and SVD2 samples separately, and for samples with
lifetimes equal to $84$ and $90\mum$. In all cases, the value of the
parameter \ctau is equal to the slope of the exponential distribution
of the selected events at the generation level within the statistical
error of the parameter \ctau.

The value of the parameter \ctau obtained from the fit to the real data
is $86.99\pm0.16\mum$. The conversion of this parameter to the value of
the $\tau$-lepton mean lifetime using the straight-line parameters of
the fit described above gives the same value: $86.99\pm0.16\mum$.
The error here is statistical.

\ifBELLE
\section{Analysis of systematic uncertainties}
\else
~
\fi

The following sources of systematic uncertainties are considered and
summarized in Table~\ref{tab:error}.

\begin{table}[htb]
\caption{Summary of systematic uncertainties\label{tab:error}}
\begin{center}\begin{tabular}{lc}
{\bf Source} & {\bf $\Delta \mean{\ctau}$  ($\mu m$)}\\ \hline
SVD alignment                           & 0.090 \\
Asymmetry fixing                        & 0.030 \\
Beam energy and ISR/FSR description     & 0.024 \\
Fit range                               & 0.020 \\
Background contribution                 & 0.010 \\ 
$\tau$-lepton mass                      & 0.009 \\ \hline
{\bf Total }                            & {\bf 0.101}
\end{tabular}\end{center}
\end{table}

A study of the influence of the SVD misalignment on the systematic shift
in the $\tau$-lepton lifetime measurement is performed in the following
way. We use $4.8$~M generated \tautau events that decay with the
$3\pi\nu_\tau - 3\pi\nu_\tau$ topology and standard Belle SVD
alignment. After all selection cuts, about $1.2$~M events remain
(compared with $1.1$~M events in the data). We shift the sensitive
elements of SVD along the $X/Y/Z$ axes by sampling from a Gaussian
function with $\sigma = 10\mum$ and rotation around these axes by
sampling from a Gaussian function with $\sigma = 0.1\mrad$. The values
of $10\mum$ and $0.1\mrad$ are obtained from the dedicated studies of
SVD alignment~\cite{SVD2}. We prepare the following decay time MC
distributions: with default alignment ($4.8$~M generated events), one
sample with misalignment according to the aforementioned shifts and
rotations ($4.8$~M generated events), several samples with misalignments
according to these shifts and rotations with fewer generated events; all
these samples have the same events at the generator level. The maximal
difference of the parameter \ctau obtained in these fits is $0.07\mum$.
This is due to the possible effect of misalignment and limited MC
statistics. We also perform global SVD shifts and rotations with respect
to the CDC by $20\mum$ and $1\mrad$, respectively. The values of
$20\mum$ and $1\mrad$ are conservative estimates from the SVD alignment
study. The variation of the \ctau parameter is within $0.06\mum$ for
these shifts. We take the value $\sqrt{0.07^2+0.06^2} = 0.09\mum$ for
the systematics due to the SVD misalignment. For an additional check of
the alignment of the tracking detectors, we divide our data sample into
two non-intersecting samples by the azimuthal ($\phi$) angle of the
momentum direction of the positive $\tau$-lepton. In the first sample
(vertical), the direction of the positive $\tau$-lepton should have
$\phi$ between $45$ and $135$ degrees or between $225$ and $315$
degrees. The second sample (horizontal) contains all the remaining
events. The obtained \ctau parameters are the same within statistical
errors, so we do not assign additional systematics due to the azimuthal
dependence of the tracking system alignment.

The systematic uncertainty due to fixing the parameter $A=2.5\cm^{-1}$
is estimated by removing the asymmetry term $(1 - A\dct)$ in the
resolution function in Eq.~\eqref{eq:res}. The difference in the
obtained lifetime, which is equal to $0.03\mum$, is taken as a
systematic uncertainty.

For the estimation of the accuracy of the initial and final state
radiation description by the KKMC generator, we analyze the
distributions of $M(\mumu)c^2 - 2\Ebeam$ for $\ee\rightarrow\mumu$
events for the data and KKMC events passed through the full Belle
simulation and reconstruction procedure. Due to the ISR and FSR, these
distributions are asymmetric and their maxima are shifted from zero to
the left. If the KKMC description of ISR and FSR energy spectrum is
harder or softer than for the data, we would observe the MC peak
position shifted from the one in the data. The result of our comparison
of the data and MC gives the difference of peak positions between the
data and MC of $(3\pm 2)\MeV$. We take the relative error
$3\MeV/10.58\GeV = 2.8\cdot 10^{-4}$ as a combined uncertainty from the
ISR and FSR description, beam energy calibration and the calibration of
the tracking system.

The variation of the fit range within about $30\%$ of that shown in
Fig.~\ref{fg:ctr} contributes an uncertainty on \ctau of $\pm 0.02\mum$.

\begin{figure}[htb]
\vspace*{-0.5cm}
\includegraphics[width=\figwidth]{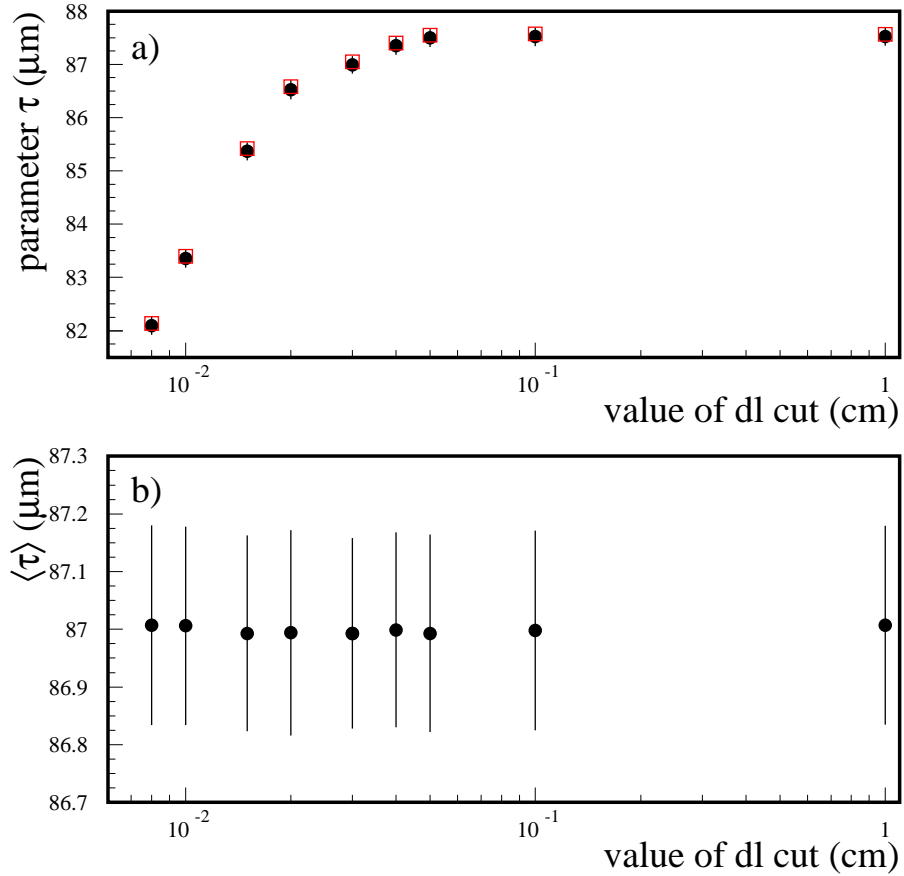}
\caption{\label{fg:dlcut}Stability when varying the value of the
$dl$-cut. a)~The dependence of the fitted parameter \ctau on the value
of the $dl$-cut for data (filled black circles) and MC (open red
squares); the errors are of the same size as the symbols.
b)~The measured value of the $\tau$-lepton lifetime as a function of the
value of the $dl$-cut; the error bar is the statistical error of the
data.}
\end{figure}

The demonstration of the stability of the obtained result to the choice
of the selection cuts is shown in Fig.~\ref{fg:dlcut}.
Figure~\ref{fg:dlcut}a shows the dependence of the fitted parameter
\ctau on the value of the cut on $dl$ for data and MC.
Figure~\ref{fg:dlcut}b shows the measured value of the $\tau$-lepton
lifetime as a function of the value of the $dl$-cut after the linear
MC-determined calibration of the parameter \ctau. One can see that this
dependence in data is very well reproduced by MC.

During the fit of the real data, the level of the background
contribution (parameter $A_{uds}$) is fixed to the nominal value
predicted by the MC in a ``nominal" fit. The contribution to the
systematic error of the \meanctau value due to the uncertainty of the
background level is tested by changing the background level in the range
of the uncertainty of the \qq{q} continuum and other backgrounds, from
$-50\%$ to $+150\%$. This range is estimated conservatively from the
control sample with looser selection criteria. The maximal variation of
the \ctau parameter is $0.01\mum$.

\ifBELLE
The relative uncertainty due to the accuracy of the $\tau$-lepton
mass~\cite{PDG} is $(0.16\MeVcc)/$ $(1776.82\MeVcc) = 9.0 \cdot 10^{-5}$.
\else
The relative uncertainty due to the accuracy of the $\tau$-lepton
mass~\cite{PDG} is $(0.16\MeVcc) / (1776.82\MeVcc) = 9.0 \cdot 10^{-5}$.
\fi

To check the stability of the result for the different periods of Belle
operation and vertex detector geometries, we repeat the analysis for
three subsamples of the data. The obtained results are consistent within
statistical errors.

\ifBELLE
\section{Lifetime difference between positive and negative
$\tau$-lep\-tons}
\else
~
\fi

The present PDG listings provide only the average lifetime of the
positive and negative $\tau$-leptons. Our measurement determines the
lifetimes for positive and negative $\tau$-leptons separately. The
difference of \meanctau for positive and negative $\tau$-leptons
obtained in the corresponding fits is $(0.07\pm 0.33)\mum$. Most of the
sources of systematic uncertainties affect the result for positive and
negative $\tau$-leptons in the same way, so their contributions to the
lifetime difference cancel. The upper limit on the relative lifetime
difference is calculated according to Ref.~\cite{FC} as
\begin{equation}
|\mean{\tau_{\tau^+}}-\mean{\tau_{\tau^-}}|/\mean{\tau_\tau}
<7.0\times10^{-3}\mathrm{\, \, at\, \, 90\%\,\, CL.}
\end{equation}
The systematic uncertainty of the lifetime difference is at least one
order of magnitude smaller than the statistical one, and is neglected.

\ifBELLE
\section{Conclusions}

\else
~
\fi

In summary, the $\tau$-lepton lifetime has been measured using the
technique of the direct decay time measurement in fully kinetically
reconstructed
$\ee\rightarrow\tautau\rightarrow 3\pi\nu_\tau\ 3\pi\nu_\tau$ events.
The obtained result for the product of the mean lifetime and speed of
light is
\begin{equation}
\mean{\ctau_\tau} = [86.99 \pm 0.16(\stat) \pm 0.10(\syst)]\mum,
\end{equation}
or in units of seconds
\begin{displaymath}
(290.17\pm0.53(\stat)\pm0.33(\syst))\cdot10^{-15}\s.
\end{displaymath}

The first measurement of the lifetime difference between $\tau^+$ and
$\tau^-$ is performed. The obtained upper limit on the relative lifetime
difference between positive and negative $\tau$-leptons is
$|\mean{\tau_{\tau^+}}-\mean{\tau_{\tau^-}}|/\mean{\tau_\tau}
< 7.0 \times 10^{-3}$ at 90\% CL.

\ifBELLE
\section{Acknowledgments}
\else
~
\fi

We thank the KEKB group for excellent operation of the
accelerator; the KEK cryogenics group for efficient solenoid
operations; and the KEK computer group, the NII, and 
PNNL/EMSL for valuable computing and SINET4 network support.  
We acknowledge support from MEXT, JSPS and Nagoya's TLPRC (Japan);
ARC and DIISR (Australia); FWF (Austria); NSFC (China); MSMT (Czechia);
CZF, DFG, and VS (Germany);
DST (India); INFN (Italy); MEST, NRF, GSDC of KISTI, and WCU (Korea); 
MNiSW and NCN (Poland); MES and RFAAE (Russia); ARRS (Slovenia);
IKERBASQUE and UPV/EHU (Spain); 
SNSF (Switzerland); NSC and MOE (Taiwan); and DOE and NSF (USA).


\begin{thebibliography}{99}
\bibitem{LU} Y.S.~Tsai,
             Phys. Rev. D {\bf 4}, 2821 (1971);
             H.B.~Thacker and J.J.~Sakurai,
             Phys. Lett. B {\bf 36}, 103 (1971).
%
\bibitem{LEPEW} S.~Schael {\it et al.}
                [ALEPH and DELPHI and L3 and OPAL and LEP Electroweak
                 Working Group Collaborations],
                arXiv:1302.3415 [hep-ex].
%
\bibitem{PDG} J. Beringer {\it et al}. (Particle Data Group),
              Phys. Rev. D {\bf 86}, 010001 (2012).
%
\bibitem{LEP} P.~Abreu {\it et al}. (DELPHI Collaboration),
              Phys. Lett. B {\bf 365}, 448 (1996); 
              G.~Alexander {\it et al}. (OPAL Collaboration),
              Phys. Lett. B {\bf 374}, 341 (1996); 
              R.~Barate {\it et al}. (ALEPH Collaboration),
              Phys. Lett. B {\bf 414}, 362 (1997); 
              M.~Acciarri {\it et al}. (L3 Collaboration),
              Phys. Lett. B {\bf 479}, 67 (2000).
%
\bibitem{Belle} A.~Abashian {\it et al.} (Belle Collab.),
                Nucl. Instr. and Meth. A {\bf 479}, 117 (2002);
                see also the detector section in
                J.~Brodzicka {\it et al.},
                Prog. Theor. Exp. Phys. (2012) 04D001.
%
\bibitem{KEKB} S.~Kurokawa and E.~Kikutani,
               Nucl. Instr. and Meth. A {\bf 499}, 1 (2003)
               and other papers included in this volume; \\
               T.Abe {\it et al.},
               Prog. Theor. Exp. Phys. (2013) 03A001
               and following articles up to 03A011.
%
\bibitem{SVD2} Z.~Natkaniec {\it et al}. (Belle SVD2 Group).
               Nucl. Instr. and Meth. A {\bf 560} 1 (2006).
%
\bibitem{V0} K.~Sumisawa {\it et al}. (Belle Collaboration),
             Phys. Rev. Lett. {\bf 95}, 061801 (2005).
%
\bibitem{taumass} K.~Belous {\it et al}. (Belle Collaboration),
                  Phys. Rev. Lett. {\bf 99}, 011801 (2007).
%
\bibitem{KKMC} S.Jadach, B.F.L.Ward and Z.W\c{a}s,
               Comp. Phys. Commun. {\bf 130}, 260 (2000).
%
\bibitem{GEANT} R.~Brun {\it et al}. GEANT 3.21.
                Report No. CERN DD/EE/84-1 (1984).
%
\bibitem{EVTGEN} D.J.~Lange,
                 Nucl. Instr. and Meth. A {\bf 462}, 152 (2001).
%
\bibitem{PYTHIA} T.~Sj\"ostrand {\it et al}.,
                 Comp. Phys. Commun. {\bf 135}, 238 (2001).
%
\bibitem{FC} 
G.J.~Feldman and R.D.~Cousins,
             Phys. Rev. D {\bf 57}, 3873 (1998); 
             J.~Conrad {\it et al.},
             Phys. Rev. D {\bf 67}, 012002 (2003).
%
\end{thebibliography}
\end{document}